# Influence of lattice distortion on the Curie temperature and spin-phonon coupling in LaMn$_{0.5}$Co$_{0.5}$O$_3$

M. Viswanathan[1+], P. S. Anil Kumar[1*], Venkata Srinu Bhadram[2], Chandrabhas Narayana[2], A. K. Bera[3], and S. M. Yusuf[3]

[1] *Department of Physics, Indian Institute of Science, Bangalore – 560012, India*

[2] *Chemistry and Physics of Materials Unit, Jawaharlal Nehru Centre for Advanced Scientific Research, Bangalore – 560064, India*

[3] *Solid State Physics Division, Bhabha Atomic Research Centre, Mumbai – 400085, India*

E – Mail : [+] viswanathan.mohandoss@yahoo.com
[*] anil@physics.iisc.ernet.in

Two distinct ferromagnetic phases of LaMn$_{0.5}$Co$_{0.5}$O$_3$ having monoclinic structure with distinct physical properties have been studied. The ferromagnetic ordering temperature $T_c$ is found to be different for both the phases. The origin of such contrasting characteristics is assigned to the changes in the distance(s) and angle(s) between Mn - O - Co resulting from distortions observed from neutron diffraction studies. Investigations on the temperature dependent Raman spectroscopy provide evidence for such structural characteristics, which affects the exchange interaction. The difference in B-site ordering which is evident from the neutron diffraction is also responsible for the difference in $T_c$. Raman scattering suggests the presence of spin-phonon coupling for both the phases around the $T_c$. Electrical transport properties of both the phases have been investigated based on the lattice distortion.

PACS numbers :   75.47.Lx, 75.47.-m, 72.80.Ga

Key words : Lattice distortion, Manganites, Magnetotransport, Raman spectroscopy, spin-phonon coupling

# I. INTRODUCTION

LaMnO$_3$ is an "A-type" antiferromagnet where the spins are aligned ferromagnetically in the *x-y* planes and the adjacent planes being stacked antiferromagnetically along the *z*-axis [1]. When transition metals are substituted in the Mn sites of LaMnO$_3$, ferromagnetism is induced [2]. Interestingly, LaMn$_{0.5}$Co$_{0.5}$O$_3$, LaMn$_{0.5}$Ni$_{0.5}$O$_3$ and LaCo$_{0.5}$Ni$_{0.5}$O$_3$ shows ferromagnetism eventhough their parent compounds LaMnO$_3$, LaNiO$_3$ and LaCoO$_3$ are not ferromagnetic [2,3]. Similarly, ferromagnetism is also induced when Mn is substituted with nonmagnetic Li$^+$ or Zn$^{2+}$ which creates Mn$^{4+}$ inducing strong ferromagnetism, while when substituted with trivalent ions like Rh$^{3+}$ or Ga$^{3+}$ causes ferromagnetism which is feeble due to the lack of Mn$^{4+}$ [4]. The origin of ferromagnetism in one such B-site doped manganite i.e. LaMn$_{1-x}$Co$_x$O$_3$ (LMCO) was initially proposed to be due to the presence of double exchange interactions [5-7] of Mn$^{3+}$ – Mn$^{4+}$, while later it was suggested that super-exchange interactions of Mn$^{3+}$ – Mn$^{3+}$ or Mn$^{4+}$ – Co$^{2+}$ could be held responsible [2,8]. LMCO undergoes a structural transition from orthorhombic to rhombohedral when $x \approx 0.6$ [2,9], where orthorhombic and rhombohedral represents the structure of the parent compounds, LaMnO$_3$ and LaCoO$_3$ respectively. Interestingly, for $x = 0.5$, the LMCO has been reported to have two ferromagnetic phases [2,9]. Initial studies revealed that the sample contained both orthorhombic and rhombohedral phases and gives rise to the two different ferromagnetic phases [2,10]. It has been shown earlier that LMCO with two different Curie temperatures ($T_c$) can be prepared, where the sample prepared at 700 $^o$C had a $T_c$ = 225 K and the other sample obtained at 1300 $^o$C had a $T_c$ = 150 K [11]. Based on neutron diffraction studies, it has been proved that the low $T_c$ phase that was misjudged as orthorhombic phase actually had monoclinic structure with the space group P2$_1$/n and a long range Mn/Co ordering [12].

Similar to B-site doped manganites, suitable A-site doping can also induce ferromagnetism in manganites. The impact of the internal pressure spanned by the A-site ion on the electronically active band formed by the overlap between *d* orbitals of B-site transition metals and *p* orbitals of oxygen, have direct control on the overlapping of orbitals, where the internal pressure is initiated by the A-site ions of different sizes [13]. Based on the earlier studies suggested by this explanation, A-site substituted LaMnO$_3$ which is ferromagnetic, showed different $T_c$, for different dopants having appropriate concentrations [14]. Although the reasons behind the changes in $T_c$ for A-site doped manganites have been proposed, the factors responsible for two distinct $T_c$ for B-site (Co) doped manganites have not been clearly understood. Based on this fact, the basic theme of our work is to identify the reasons for the occurrence of two ferromagnetic phases in LMCO.

In this paper, we evaluate the structural aspects of LMCO which reflect on its intrinsic characteristics. Similar to earlier reports from XRD, our results also support the high $T_c$ phase to have rhombohedral structure. But interestingly, from neutron diffraction (ND) we firmly come to a conclusion that the high $T_c$ phase also has a monoclinic structure. Neutron diffraction of the high $T_c$ phase is reported for the first time. Investigations based on neutron diffraction reveal that the B-site ordering to be higher for the high $T_c$ sample. We substantiate that each phase having the same crystal structure has distinct inter cationic distance(s) ($d_{TM-TM}$), inter cationic angle(s) ($\theta_{TM-O-TM}$) and B-site ordering, being responsible for the difference in $T_c$. The two phases of LMCO with distinct $d_{TM-TM}$, $\theta_{TM-O-TM}$ and B-site ordering have been investigated in which the distortion aided changes in transport and magnetic properties are studied. Raman scattering is an extremely sensitive physical tool to examine lattice distortions, spin-phonon coupling etc., and has been extensively used in manganite systems to get the insight into its physical properties. In this paper, we have used Raman spectroscopy to corroborate the lattice distortions observed from the results arrived at from Rietveld analysis to provide conclusive evidence on the structural aspects of LMCO. Based on the temperature dependent Raman scattering studies on both the phases, we verify that there is strong spin-phonon coupling around $T_c$ for LMCO for both the structures. In an earlier report [15] a sharp discontinuity in Raman frequency at 150 K, coincident with magnetic transition was assigned to spin-lattice interaction, consistent with our observation of strong spin-phonon coupling in this system.

## II. EXPERIMENTAL DETAILS

The Low $T_c$ sample was prepared by solid state reaction with $La_2O_3$, $MnO_2$ and $Co_3O_4$ as the precursors. The samples were heated to 1370°C and furnace cooled with several intermediate grindings. Preparation of this sample by quenching in Liq. $N_2$ as reported earlier [10] was avoided since quenching induced strains might result in peak shifts in XRD/ND leading to ambiguous conclusions during Rietveld refinement. The High $T_c$ sample was prepared by the glycine-nitrate method. The prepared powder was then heated to 790°C and furnace cooled with a few intermediate grindings. The oxygen stoichiometry was examined for both the samples by iodometric titration and was experimentally verified to be nearly stoichiometric within the experimental limits. XRD was taken using a Bruker D8 Advance diffractometer with Cu Kα radiation. Neutron diffraction experiments were performed at room temperature using the five linear position sensitive detectors based neutron powder diffractometer (λ = 1.2443 A°) at Dhruva reactor, Bhabha Atomic Research Centre, India. Raman experiments were carried out using a custom built Raman spectrometer [16]. The LASER excitation used was 532 nm from a frequency doubled solid state Nd-YAG laser with a lower power of 8 mW on the sample. The measurements were carried out in a back scattering geometry from 80 K to 300 K ± 1K.

## III. RESULTS AND DISCUSSION

The crystallographic details of the powder samples were analysed by Rietveld refinement of X-ray diffraction and are shown in figure 1. Both the samples were found to be single phase. The Low $T_c$ sample could be indexed as monoclinic (LTcM) with space group identified as P2$_1$/n ($\chi^2$ = 3.756; $R_{wp}$ = 2.79 %), while the High $T_c$ sample could be indexed as rhombohedral (HTcR) with R3C space group ($\chi^2$ = 3.367; $R_{wp}$ = 2.70 %). However it has to be noted that the high $T_c$ sample could also be fitted for P2$_1$/n ($\chi^2$ = 4.101; $R_{wp}$ = 2.98 %). In the case of LTcM, a reduction in symmetry is clearly evident from the splitting of the main peak as shown in the inset of figure 1(a). The values of lattice parameters, bond lengths, and bond angles are mentioned in Table 1. We observe that $d_{TM-TM}$ and $\theta_{TM-O-TM}$ are different for both the samples. $\theta_{TM-O-TM}$ ranges from 156.85$^0$ to 162.00$^0$ for the

LTcM, while it is 167.84$^0$ for the HTcR. The average bond length in LTcM, $\langle d_{Mn-O} \rangle$ = 2.024 Å, while it is shorter in case of HTcR where $\langle d_{Mn/Co-O} \rangle$ = 1.950 Å.

We rely more on neutron diffraction in contrast to X-ray diffraction, since the methods based on the later are less sensitive to low Z elements such as oxygen. Thus, using XRD in locating the atomic positions of oxygen atoms becomes inaccurate leading to improper conclusions. The monoclinic structure for LTcM is well supported by the present ($\chi^2$ = 41.7; $R_{Bragg}$ = 6.57 %) and the earlier results [12] based on ND. While in case of HTcR (for which no investigations have previously been done using ND), all observed Bragg peaks in the neutron diffraction pattern could not be indexed with rhombohedral space group R3C, suggesting a lower symmetry for this compound. All observed Bragg peaks can be indexed with the monoclinic space group P2$_1$/$n$, as observed for LTcM, This gives a better agreement between the observed and the calculated (using Rietveld refinement) diffraction patterns ($\chi^2$ = 20.4; $R_{Bragg}$ = 5.9 %). Thus, it is clearly evident that although it is possible to fit XRD results to R3C, ND clearly shows a few peaks being un-indexed which were not even detected by XRD. From the Rietveld analysis, $U_{iso}$ values of oxygen were found to be 0.96(3)/1.15(6)/0.95(3) and 0.84(1)/1.15(3)/0.95(4) for LTcM and HTcM respectively. These values of $U_{iso}$ of oxygen are comparable to other perovskites [17,18]. Further discussions on high Tc sample with monoclinic structure will be denoted as HTcM instead of HTcR. It should be noted that the occupancy of all the elements as arrived from the Rietveld refinement of XRD and neutron diffraction shows that the composition of both the phases are the same (within 1% in case of XRD and less than 0.5% in the case of neutron diffraction). Using both the techniques it was also observed that there is a minor deficiency at the B-site for both the samples and both the samples and they are compositionally identical to each other. Figure 2 shows the ND results for both LTcM and HTcM. The lattice parameters and bond details arrived from ND are given in table 2. We observe that $d_{TM-TM}$ and $\theta_{TM-O-TM}$ are different for both the samples. $\theta_{TM-O-TM}$ ranges from 155.2$^0$ to 165.3$^0$ and 159.4$^0$ to 161.0$^0$ for LTcM and HTcM samples, respectively. As $\theta_{TM-O-TM}$, tends to 180$^0$ the lattice distortion gets reduced. This clearly elucidates that the lattice distortion in LTcM is lower than that of HTcM. The average bond length in LTcM, $\langle d_{TM-O} \rangle$ = 1.975 Å, while it is longer in case of HTcM where $\langle d_{TM-O} \rangle$ = 1.980 Å. These observations of different $\theta_{TM-O-TM}$ and $d_{TM-O}$ are directly correlated to the $T_c$ of the LMCO as discussed below.

In the earlier studies [11] it has been reported that orthorhombic phase ($T_c$ ~ 150 K) exhibiting an ordered moment ($\mu$) of 4.01 $\mu_B$ could be assigned either to Mn$^{3+}$ high spin (HS, t$_{2g}^3$ e$_g^1$) – Co$^{3+}$ intermediate spin (IS, t$_{2g}^5$ e$_g^1$) or Mn$^{4+}$ – Co$^{2+}$ (HS, t$_{2g}^5$ e$_g^2$) state, while in the case of rhombohedral phase ($T_c$ ~ 225 K) having $\mu$ of 3.52 $\mu_B$ the spin state was assigned to Mn$^{3+}$ (HS) – Co$^{3+}$ (LS, t$_{2g}^6$ e$_g^0$). Later it was shown to be of Mn$^{4+}$ – Co$^{2+}$ (HS) character for both the phases while the monoclinic phase contained Co$^{3+}$ (LS) [19].

In the present work, in order to understand the magnetic properties, the temperature dependent DC and AC susceptibility measurements were done at a constant magnetic field of $H_{DC}$=100 Oe and $H_{AC}$=170 mOe with a frequency of 420 Hz respectively. FC and ZFC magnetisation are shown in figures 3(a) and (b) for LTcM and HTcM respectively. The $T_c$ for LTcM and HTcM are found to be 123 K and 232 K respectively exhibiting a "Brillouin like" feature which is a characteristic signature of ferromagnetic materials. When compared to DC magnetic studies (such as SQUID/VSM/AGM), the AC susceptibility is a sensitive technique to probe $T_c$ at very low magnetic fields, which could easily point out the presence of more than one magnetic phase. These measurements revealed that the LTcM and the HTcM samples had ferromagnetic transitions at 138 K and 243 K respectively as shown in figures. 4(a) and (b). We have not observed any frequency dependent (75 to 1000 Hz) changes in $T_c$ which suggest the presence of ferromagnetic ordering. Though the single phase nature was observed by XRD/ND in LTcM, inset of figure 4a representing $\chi'^{-1}$ shows a small kink around 225 K which may be due to the presence of an extremely small residual high $T_c$ phase.

In order to understand the change in $T_c$, we have used the analogy from A-site doped systems. In general, $T_c$ depends on total angular moment ($J$), nearest number of neighbours (n) and exchange energy ($J_{ex}$) given by $T_c = [2nJ_{ex} J(J+1)]/3k_B$. A variation in $T_c$ was observed in A-site doped LaMnO$_3$ ([i]Pr$_{0.7}$Ca$_{0.3}$MnO$_3$, [ii] La$_{0.525}$Pr$_{0.175}$Ca$_{0.3}$MnO$_3$) [14], where the radii of the A-site ions had influenced internal pressure causing changes in the distances and angles between transition metals (i.e. $\Delta d_{TM-TM}$ ~ 0.013 Å; $\Delta\theta_{TM-O-TM}$ ~ 1.9° /3.6°). The change in $T_c$ i.e. $\Delta T_c$ in this case is ~ 100 K. The transfer integral t$_{ij}$ is found to control the double exchange interactions in manganites [20], a direct outcome of $\Delta\theta_{TM-O-TM}$. In the case of LMCO both the phases have identical n. Hence we attribute that the $\Delta d_{TM-TM}$ and $\Delta\theta_{TM-O-TM}$ are responsible for $\Delta T_c$ ~ 105 K. This scenario could be accepted by comparing the structures of LTcM and HTcM, where $\Delta d_{TM-TM} \approx$ 0.0137 Å / 0.0048 Å / 0.0048 Å; $\Delta\theta_{TM-O-TM}$ ~ 1.2° /4.3° /4.3° for three different atomic positions of oxygen atoms respectively for LTcM when compared to HTcM.

On the crystallographic perspective, the tolerance factor, $t = (d_{A-O})/\sqrt{2}(d_{B-O}) = 1$ for cubic perovskites with $\theta_{TM-O-TM} = 180°$, where $d_{A-O}$ is the bond length of A-site atom and oxygen, and $d_{B-O}$ corresponds to the bond length of B-site atom and oxygen. When $t < 1$, octahedra get tilted at the expense of $\theta_{TM-O-TM}$, initiating geometrical distortions due to mismatch of A-site ions in A-site substituted manganites, while in our case it is due to $d_{B-O}$. In our case of B-site doping, a tilting of octahedra is observed where the average tolerance factor $\langle t \rangle$ is, $\langle t$ (LTcM)$\rangle$ = 0.9473; $\langle t$ (HTcM)$\rangle$ = 0.9470. Correlation between higher $\theta_{TM-O-TM}$ and higher $\langle t \rangle$ is justifiable for LTcM which less distortion.

Apart from the lattice effects, B-site ordering was also examined from ND. The degree of ordering is estimated from the intensity of (101) and (011) Bragg peaks. Figure 5(a). shows the calculated diffraction patterns for different degree of ordering (0 to 100 %) of Mn and Co ions at the 2$d$ and 2$c$ sites. The integrated intensity of (101) and (011) peaks is found to be higher for better ordering. From experimental data we find that the integrated intensity of HTcM is higher than LTcM as shown in figure 5(b). From the experiments, it is found that HTcM has a very high ordering of 81.2 % when compared to LTcM with 47.6 %, while the same is shown in the inset of figure 5(c) in comparison with the calculated values. In perovskite systems, it has been observed that B-site ordering can affect magnetic properties like Curie temperature [21]. In general, the difference in charge between the cations in the B-site determines better cationic ordering [22]. Earlier it has been shown that the degree of ordering in perovskite systems depend on synthesis parameters such as time [22], temperature [23] (i.e better ordering for more the time/temperature of heat treatment). This convention is not true in case of LMCO. Better B-site ordering is found if the difference in charge between the B-site cations is more than two [21]. Using XAS, it was reported that the residual existence of $Mn^{3+}$ in the low $T_c$ sample was more than the high $T_c$ sample [19]. The presence of $Mn^{3+}$ reduces the effective difference in charge of B-site cations less than two and thus deteriorates the ordering. As the presence of $Mn^{3+}$ is more in LTcM when compared to HTcM it has a direct effect on the ordering and is well correlated with the results obtained from ND. Jahn Teller distortion exhibited by $Mn^{3+}$ can also have an influence in the electronic properties.

In order to understand the lattice distortion in detail, both LTcM and HTcM were examined by Raman scattering, while its temperature dependent investigations can be used to examine the spin-phonon coupling. From the lattice dynamical calculations (LDC), LMCO with monoclinic and rhombohedral structures were predicted to have 24 and 8 Raman modes respectively [24]. It has been shown that LMCO has similar TM – O octahedral vibrations in comparison to $LaMnO_3$ (Pbnm) and octahedral fluoride complexes and their predominant peaks at 490 and 645 $cm^{-1}$ were assigned to Antistretching ($\omega_A$) and Stretching modes ($\omega_S$) respectively [25]. LDC confirms that the peak at 697 $cm^{-1}$ is assigned to a stretching mode ("breathing"), while the ones at 490 $cm^{-1}$ is of a mixed type ($\omega_{A,B}$ i.e. antistretching and bending). Although LDC suggests that $\omega_S$ ($P2_1/n$) ~ 697 $cm^{-1}$, experimentally it was verified to be 645 $cm^{-1}$ [24].

Figure 6(a) and (b) shows the temperature dependent Raman spectra for LTcM and HTcM respectively. The intensity of the rotational mode ($\omega_R$) around 250 $cm^{-1}$ (not shown) is weak and is not included in the discussion. From figure 6 we find that $\omega_{A,B}$(LTcM) > $\omega_{A,B}$(HTcM) ($\Delta\omega_{A,B}$ ~25 $cm^{-1}$), whereas $\Delta\omega_S$ is a not more than 5 $cm^{-1}$ for the entire temperature range. Such closeness of $\omega_S$ is observed between LTcM and HTcM, since $\langle d_{TM-O}(LTcM) \rangle$ ~ 0.9974 $\langle d_{TM-O}(HTcM) \rangle$ (from Table II) from which we infer that $\omega_S$ is similar as the way the bond lengths are and also $\theta_{O-TM-O}$ is 180.0° for both LTcM and HTcM ruling out the possibility of the octahedra itself being self-distorted. The dependence of $\omega_{A,B}$ on temperature is plotted for LTcM and HTcM in figure 7(a), which shows that $\omega_{A,B}$(LTcM) > $\omega_{A,B}$(HTcM) for the entire temperature range. In the case of $\omega_{A,B}$(HTcM), the initial hardening effect is due to $\omega_{latt}$ i.e $d_{TM-O}$ as $\phi$ slightly varies and subsequently relaxes to an equilibrium position ensuring $\omega_{A,B}$ becomes almost invariable below a certain temperature. Such kind of initial hardening is found to be less featured in LTcM.

In the case of manganese oxides, the $MnO_6$ octahedra when rotated along the [111] axis give rise to a rhombohedral symmetry, and when rotated along the [110] axis results in an orthorhombic symmetry. Such rotations reduces the $\theta_{Mn-O-Mn}$ from ideal 180° to (180° – $\phi$), for which $\phi$(R3C) < $\phi$(Pbnm/Pnma) [13], where $\phi$ is the tilt angle of the octahedra. The octahedra are tilted in such a way that $\langle d_{La-O} \rangle$ has optimal bond length affecting the spring constant of the vibrating system. In our case, $\langle d_{La-O} \rangle$ = 2.646 A° and 2.652 A° for LTcM and HTcM respectively. In general, the tilt angle of the octahedra and $\langle d_{La-O}\rangle$, controls the rotational and bending modes respectively. When compared to HTcM, $\langle d_{La-O}\rangle$ being shorter for LTcM explains the higher frequency of vibrations of the bending mode. Although it is true that for higher values of φ (i.e more distortion) $\omega_{A,B}$ should be high, it is not observed on our case as both $\langle d_{La-O}\rangle$ and $\langle d_{TM-O}\rangle$ are higher for HTcM and thus reduce the bending frequency when compared to LTcM. Comparisons of $\langle t\rangle$, $\omega_{A,B}$ and φ at 300 K are shown in figure 7(b). A similar Raman shift dependence on $\langle t\rangle$ was evaluated for A-site doped manganites, where $\langle t\rangle$ is dependent on the A-site ion [26] which is due to the internal pressure spanned by the A-site ion [14]. While such A-site doped manganites with $\langle t\rangle > 0.925$ had rhombohedral structure, our case of B-site (Co) doped $LaMnO_3$, retains $P2_1/n$ for higher values of $\langle t\rangle$ i.e. $\langle t\rangle = 0.9473/ 0.9470$.

Figures 8(a) and (b) shows the temperature dependence of the stretching mode ($\omega_S$) for LTcM and HTcM. Comparing $\omega_S$ and Rietveld analysis (both at 300 K), $\omega_S$(LTcM) and $\omega_S$(HTcM) are very close (within the error values) since there is no difference in $\theta_{O-TM-O}$. Also the values of $d_{TM-O}$ being very close to each other could be the reason for smaller magnitude of $\Delta\omega_S$. The Phonon frequencies has an effect on the FM ordering below $T_c$ because in case of magnetic materials, $\Delta\omega(T)$ depends on lattice expansion/contraction ($\Delta\omega_{latt}$), anharmonic scattering ($\Delta\omega_{anh}$), phonon renormalisation of the electronic states that occurs near the spin ordering temperature due to electron phonon coupling ($\Delta\omega_{ren}$), and the spin-phonon contribution which is caused by the modulation of exchange integral by lattice vibrations ($\Delta\omega_{s-ph}$) [27]. $\Delta\omega_{ren}$, can be ignored as the carrier concentration is low.

In order to understand both the phases individually, we have attempted to study the spin-phonon coupling for both LTcM and HTcM individually. In both the cases of LTcM and HTcM, $\omega_S$ deviates from the regular dependence [28, 29] of $\omega_{anh}(T) = \omega_o - C(1+(2/e^{(h\omega_0/2k_BT)}-1))$ for $T < T_c$, clearly indicating the nature of spin-phonon coupling as shown in figure 8. In the case of LTcM, $\omega_S$ increases (hardens) as the temperature is lowered. This hardening is due to anharmonicity ($\omega_{anh}$). Such kind of anharmonic nature is observed with mild deviation around 230 K, which is a clear signature of $\Delta\omega_{s-ph}$ for the residual HTcM as inferred from $\chi'^{-1}$. Such effects are also observed from the FWHM of $\omega_s$, which is related to phonon lifetime where there is a slope change around 230 K due to $\Delta\omega_{s-ph}$ (HTcM). Below 230 K the behaviour of $\omega_S$ is driven by the LTcM phase. Interestingly, around 175 K we observe an anomalous behaviour in $\omega_S$. This is due to two competing factors namely, strong spin-phonon coupling of HTcM and approaching transformation in LTcM. Hence this softening of $\omega_S$ below 175 K is due to increasing influence of the HTcM phase. Below 125 K, the LTcM develops a strong spin-phonon coupling shown by the FWHM (see figure. 8(a)).

In the case of HTcM, anharmonic nature is observed till the temperature is lowered upto 235K (~ $T_c$ of HTcM phase). Below 235 K, the behaviour of $\omega_S$ is anomalous. It shows two distinct changes around 235 K and 125 K. Incidentally, these temperature corresponds with $T_c$ of HTcM and LTcM respectively. It is interesting to note that Raman scattering shows evidence of the presence of LTcM in the HTcM phase, not observed by other methods. In order to explain this anomalous behaviour we have looked at the FWHM of $\omega_S$, which is related to phonon lifetime. The inset of figure 8(b) shows anomalous behaviour in phonon lifetime around 240 K and 125 K. On decreasing temperature, one expects the FWHM to decrease. This deviation from the expected behaviour suggests the increase in phonon lifetime, which is consistent with the decrease in phonon energy. This is due to a strong spin-phonon coupling. The presence of LTcM is the reason for the dual effect. These observations are clear indications of the residual presence of one phase in the other. Residual phase can be detected from its $\Delta\omega_{s-ph}$ which could be observed from inelastic scattering, as it is a local probing technique unlike magnetic methods. For a highly ordered system the phonon life time is higher i.e. the FWHM is smaller. From the insets of figure 8 it is clear that the phonon life-time is higher for HTcM as it has better B-site ordering. These results are well supportive to the conclusions arrived from ND.

We have also investigated the transport characteristics of LMCO, individually for both the phases as shown in figure 9. For the entire temperature regime, the samples exhibit an insulating nature irrespective of the degree of distortion, with resistivity of ρ(HTcM) > ρ(LTcM). In the case of LTcM, for $T > T_c$, (i.e. 155 to 300 K, the paramagnetic region) Efros-Shklovskii (ES) VRH [30] as given by $\rho = \rho_0 \exp(T_0/T)^{\eta}$ was observed (with $\eta = 0.5$), while we were unable to fit the data to many of the conduction mechanisms such as VRH, polaron hopping and thermally activated transport for the HTcM. This could be understood if one takes a closer look at the bending modes observed by Raman scattering. In the case of the paramagnetic region of HTcM, $\omega_{A,B}$ exhibits a continuous change, which might be a factor which doesn't permit the transport to follow any possible conduction mechanism, while in case of LTcM the bending modes are almost same i.e. ϕ is independent of temperature. To have a better understanding of HTcM, one should conduct high temperature transport studies. The ρ(HTcM) > ρ(LTcM) can be understood based on the difficulty faced by an electron to hop from one TM to another due to $\theta_{TM-O-TM}$ being lower indicating a greater deviation from $180^o$ i.e. $\Delta\theta_{TM-O-TM}$ ~ $1.2^o$ /$4.3^o$ /$4.3^o$.

Magnetoresistance (MR) upto 11 Tesla has also been measured for LMCO and both the samples exhibited negative MR. LTcM and HTcM samples exhibited a maximum change in resistance of ~ 51 % and ~ 31 % respectively at 125 K as shown in the inset of figure 9. This reduction in resistance in presence of the magnetic field could be due to the augmented probability of electron hopping due to the alignment of magnetic moment of transition metal ions in the presence of a magnetic field.

**IV. CONCLUSION**

LMCO with two different ferromagnetic transition temperatures have been studied by AC susceptibility and DC Magnetisation. The difference in the $T_c$ for LaMn$_{0.5}$Co$_{0.5}$O$_3$ has been evaluated and it could be understood on the basis of the degree of distortion present in the two phases. Previous studies on LMCO suggested that it can have both R3C and Pbnm structures. While later investigations proved that the low $T_c$ phase is P2$_1$/n and not *Pbnm*, there was no any conclusive evidence for the structure of high $T_c$ phase. Our Rietveld analysis of neutron diffraction and Raman spectroscopy confirms that the high $T_c$ phase is also P2$_1$/n. Neutron diffraction studies also confirm a better B-site ordering in high $T_c$ sample and is well supported by the higher phonon life-time observed by Raman spectroscopy. The observed difference in $T_c$ for the two phases is due to the lattice distortion influenced by the tilting of octahedra pertaining to the Mn/Co ordering directly affects the exchange energy. The temperature dependent Raman spectra confirm that the deviation of anharmonicity around $T_c$ indicates the presence of strong spin-phonon coupling in both the phases of LMCO. We have also evaluated the magnetotransport characteristics of both the phases.

## ACKNOWLEDGEMENT

We thank the National facility for Low temperature and High magnetic fields, IISc.

| Sample | Space group | a (Å) | b (Å) | c (Å) | Angle (θ in degrees) |
|---|---|---|---|---|---|
| LTcM | $P2_1/n$ | 5.523 | 5.480 | 7.763 | $\alpha = \gamma = 90$; $\beta = 90.024$ |
| HTcR | $R\bar{3}C$ | 5.496 | 5.496 | 13.372 | $\alpha = \beta = 90$; $\gamma = 120$ |

| Sample | Angle (degrees) θ | Bond length (Å) $d_{TM-O}$ | Bond length (Å) $d_{TM-TM}$ |
|---|---|---|---|
| LTcM | O1 – TM – O1 = 180.00<br>O2 – TM – O2 = 180.00<br>O3 – Co – O3 = 179.97<br>O3 – Mn – O3 = 179.98<br>Co – O1 – Mn = 162.00<br>Co – O2 – Mn = 156.65<br>Co – O3 – Mn = 158.29 | Co – O1 = 1.931<br>Co – O2 = 1.926<br>Co – O3 = 1.934<br><br>Mn – O1 = 1.999<br>Mn – O2 = 2.046<br>Mn – O3 = 2.027 | Mn – Co = 3.882 (for O1)<br><br>Mn – Co = 3.890 (for O2)<br><br>Mn – Co = 3.890 (for O3) |
| HTcR | O – TM – O = 179.98<br>TM – O – TM = 167.84 | TM – O = 1.950 | Mn – Co = 3.878 |

Table 1. Results from the Rietveld refinement are given. The lattice parameters, bond lengths and the angles associated with the transition metals (TM) and oxygen is presented.

| Sample | Space group | a (Å) | b (Å) | c (Å) | Angle (θ in degrees) |
|---|---|---|---|---|---|
| LTcM | $P2_1/n$ | 5.5311(2) | 5.4818(1) | 7.7791(1) | $\alpha = \gamma = 90$; $\beta = 89.911(5)$ |
| HTcM | $P2_1/n$ | 5.5104(1) | 5.5162(1) | 7.8063(2) | $\alpha = \gamma = 90$; $\beta = 89.352(3)$ |

| Sample | Angle (degrees) θ | Site | Bond length (Å) $d_{TM-O}$ | Bond length (Å) $d_{TM-TM}$ |
|---|---|---|---|---|
| LTcM | O1 – TM – O1 = 180.0(1)<br>O2 – TM – O2 = 180.0(2)<br>O3 – TM – O3 = 180.0(2) | 2d | TM – O1 = 2.000(3)<br>TM – O2 = 1.935(4)<br>TM – O3 = 1.930(5) | Mn – Co = 3.8895(8) (for O1)<br><br>Mn – Co = 3.8937(4) (for O2)<br><br>Mn – Co = 3.8937(4) (for O3) |
|  | Co – O1 – Mn = 160.6(3)<br>Co – O2 – Mn = 165.3(6)<br>Co – O3 – Mn = 155.2(6) | 2c | TM – O1 = 1.940(3)<br>TM – O2 = 1.991(1)<br>TM – O3 = 2.055(3) |  |
| HTcM | O1 – TM – O1 = 180.0(1)<br>O2 – TM – O2 = 180.0(2)<br>O3 – TM – O3 = 180.0(2) | 2d | TM – O1 = 1.901(3)<br>TM – O2 = 1.940(7)<br>TM – O3 = 1.960(7) | Mn – Co = 3.9032(9) (for O1)<br><br>Mn – Co = 3.8985(5) (for O2)<br><br>Mn – Co = 3.8985(5) (for O3) |
|  | Co – O1 – Mn = 159.4(5)<br>Co – O2 – Mn = 161.0(7)<br>Co – O3 – Mn = 159.8(7) | 2c | TM – O1 = 2.066(3)<br>TM – O2 = 2.012(7)<br>TM – O3 = 2.002(7) |  |

Table 2. Results from the Rietveld refinement of neutron diffraction.

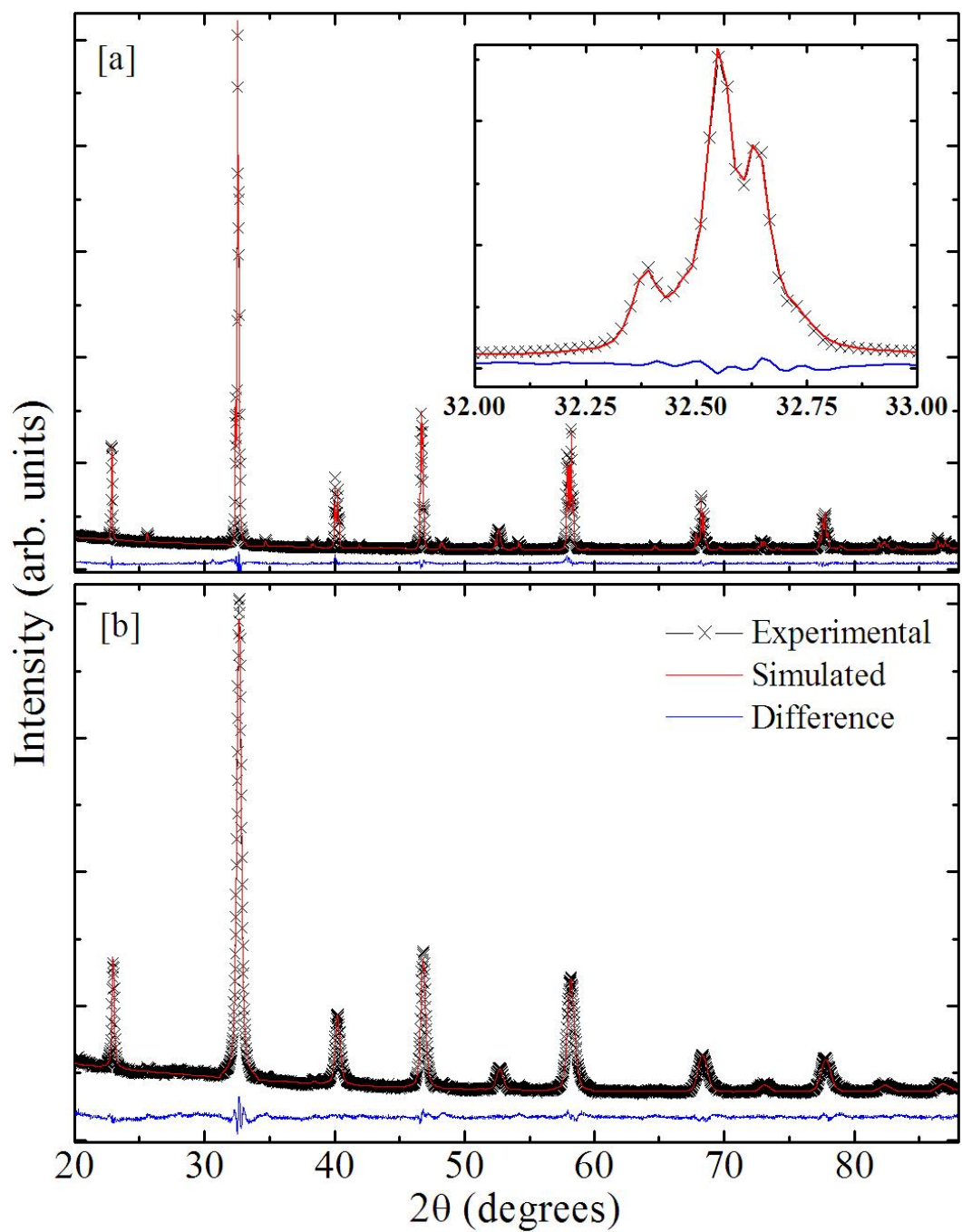

Figure 1. X-ray diffraction of LaMn$_{0.5}$Co$_{0.5}$O$_3$ [a] LTcM and [b] HTcR. Inset of figure 1(a) shows the splitting being a characteristic of monoclinic.

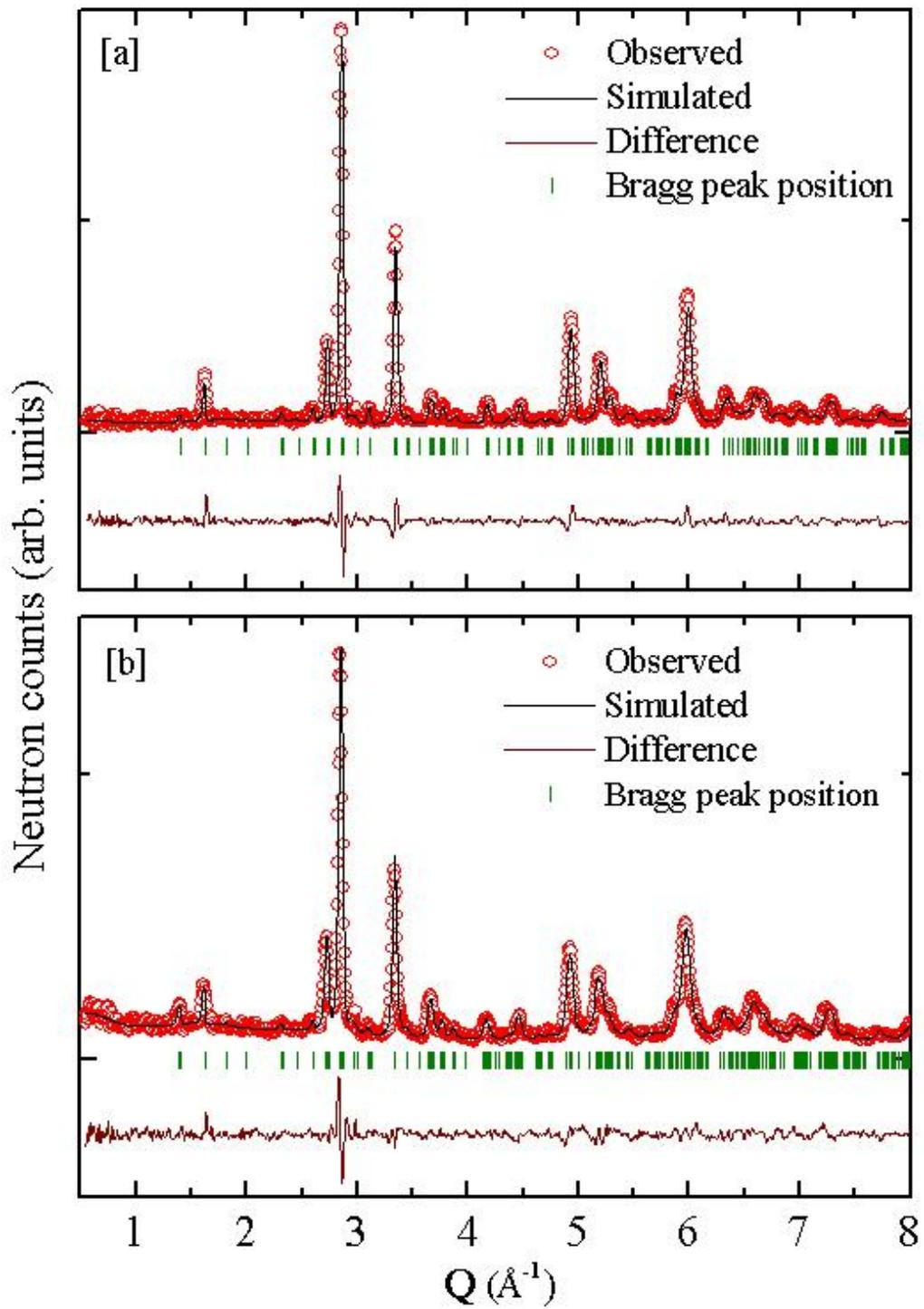

Figure 2. Neutron diffraction of LaMn$_{0.5}$Co$_{0.5}$O$_3$ [a] LTcM and [b] HTcM.

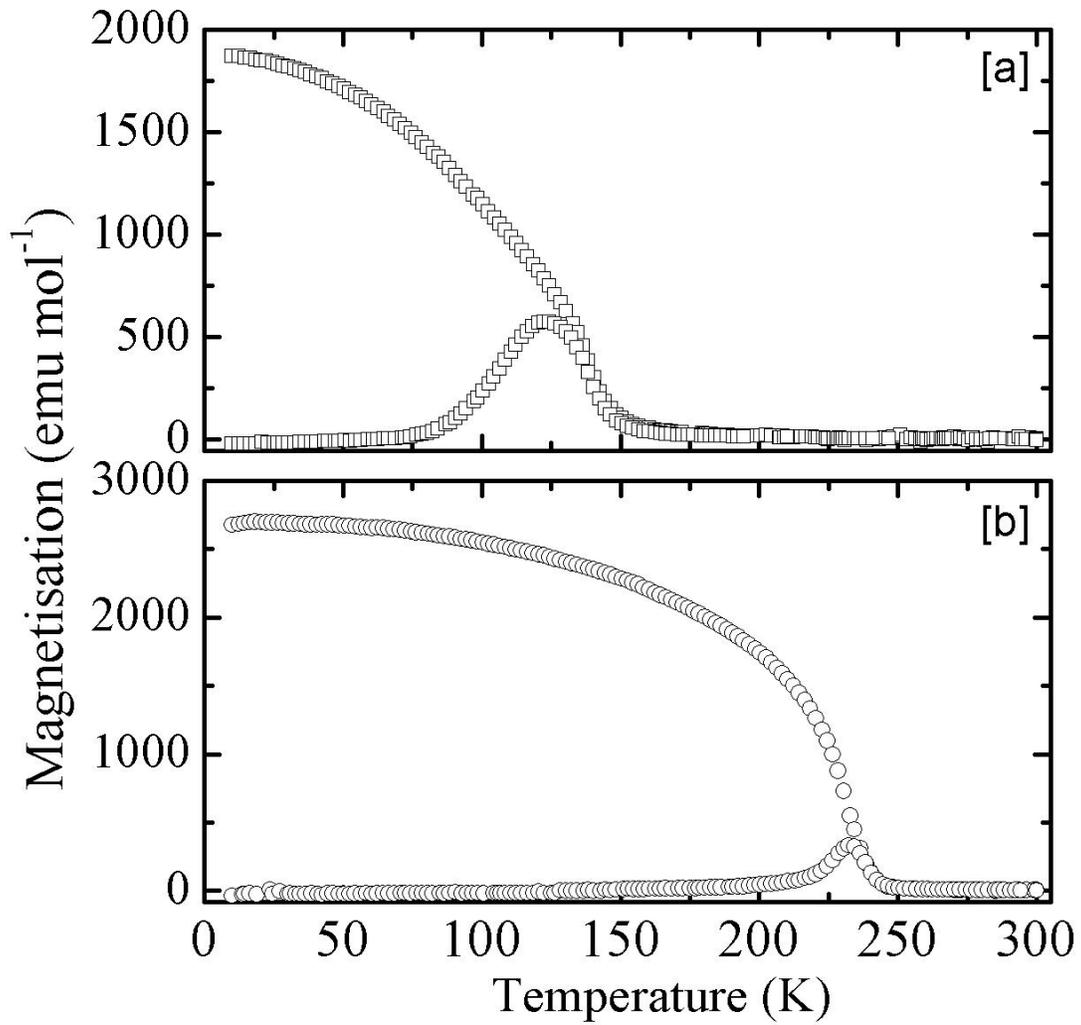

Figure 3. DC magnetisation at $H_{DC}$ = 100 Oe, showing the "Brillouin like" characteristic feature of ferromagnetic $LaMn_{0.5}Co_{0.5}O_3$. [a] LTcM and [b] HTcM.

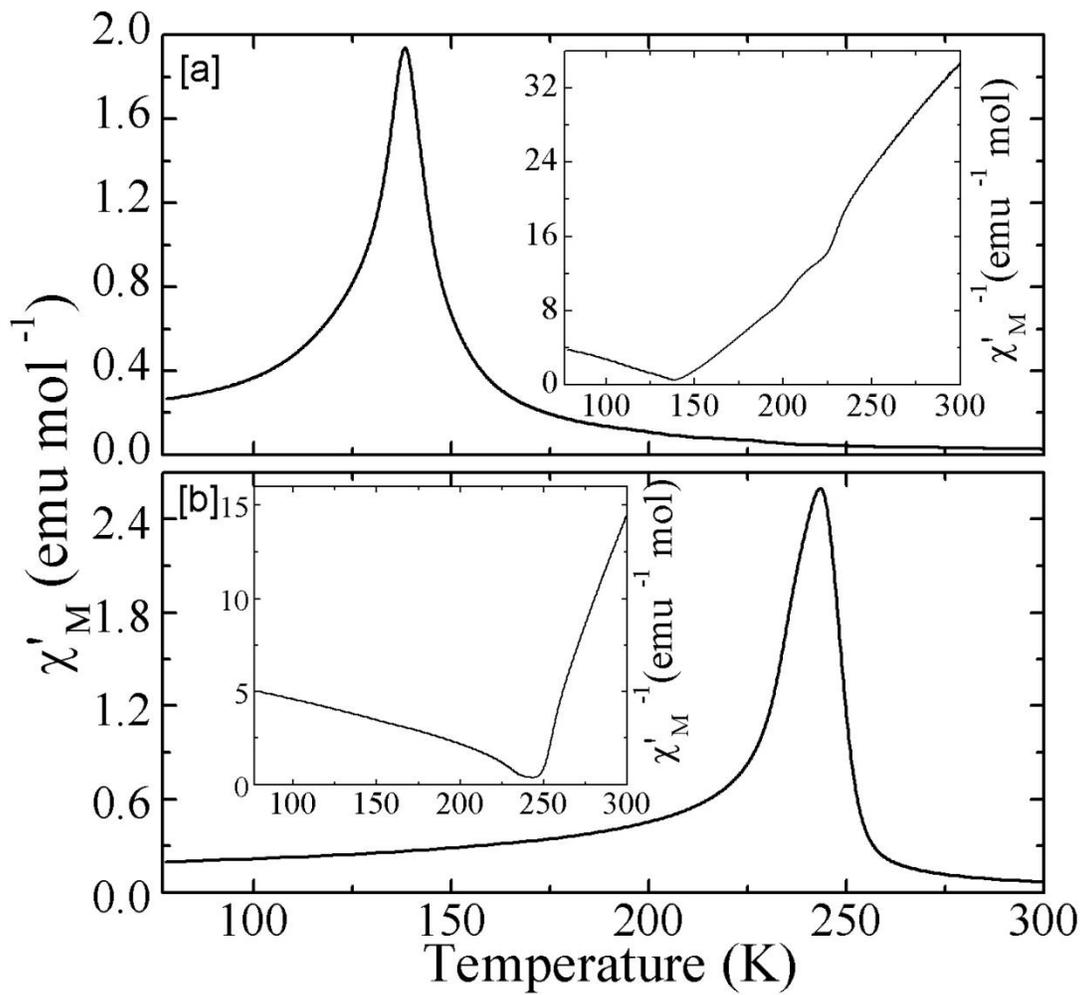

Figure 4. In-phase component of Molar AC Susceptibility at 170 mOe and a frequency of 420 Hz for [a] LTcM and [b] HTcM phases. Insets show their respective $\chi'^{-1}$.

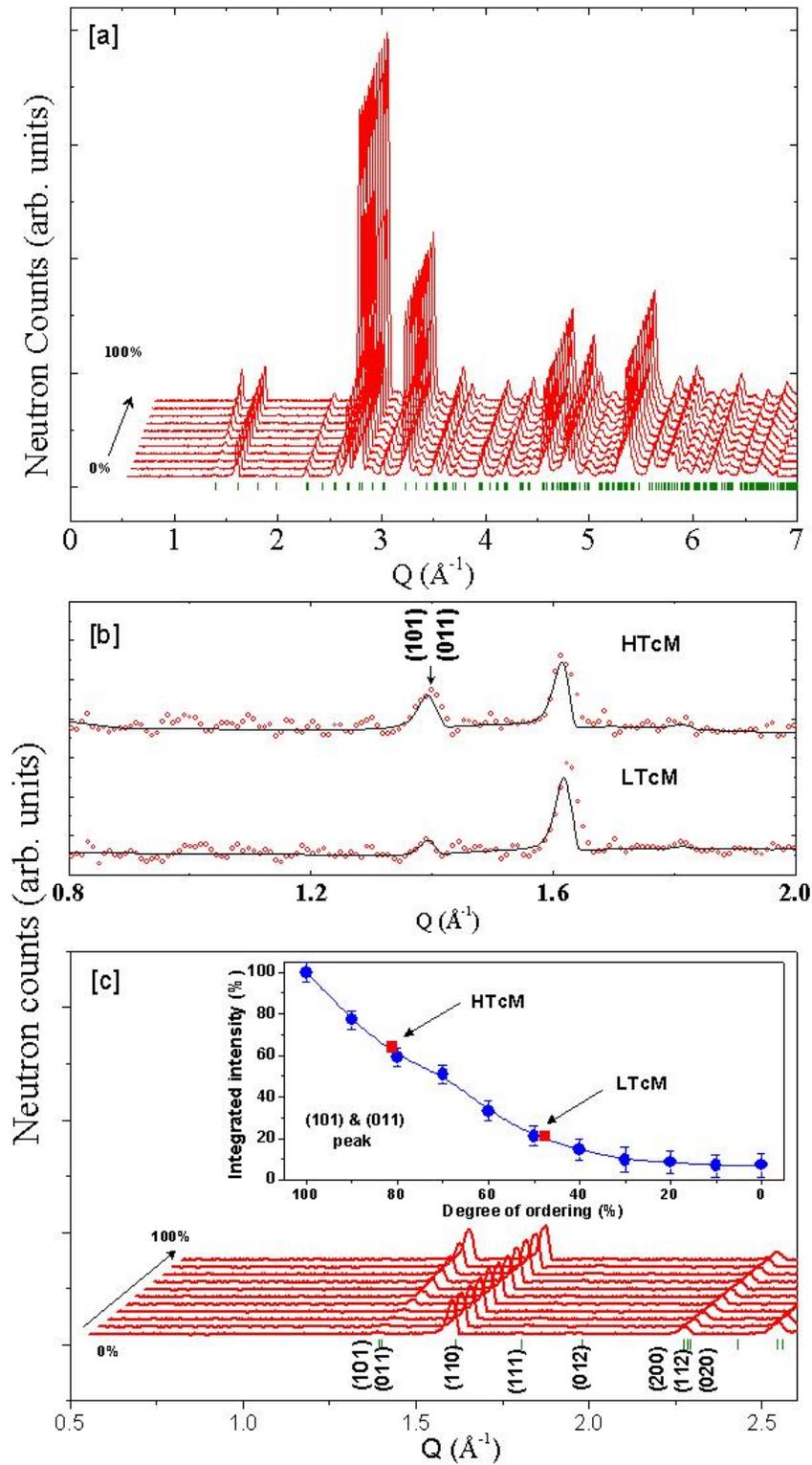

Figure 5. (a) Calculated diffraction patterns for various B-site ordering. (b) Expanded region of (101) and (011) peaks which represents the degree of B-site ordering. (c) Close view of the low Q region of the calculated patterns and an inset showing the integrated intensity as a function of B-site ordering, where experimental values of LTcM and HTcM are denoted.

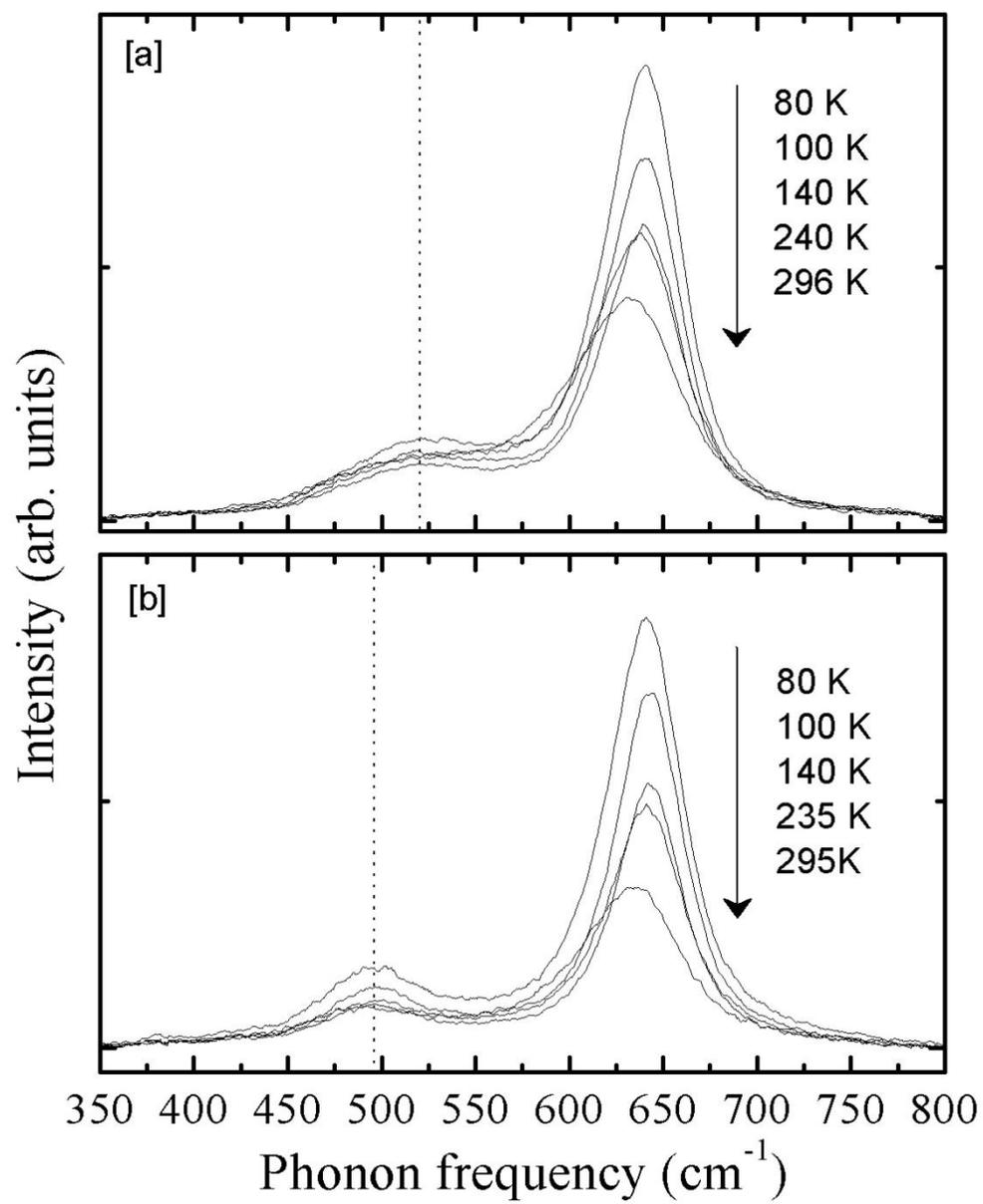

Figure 6. Temperature dependent Raman spectra for [a] LTcM and [b] HTcM.

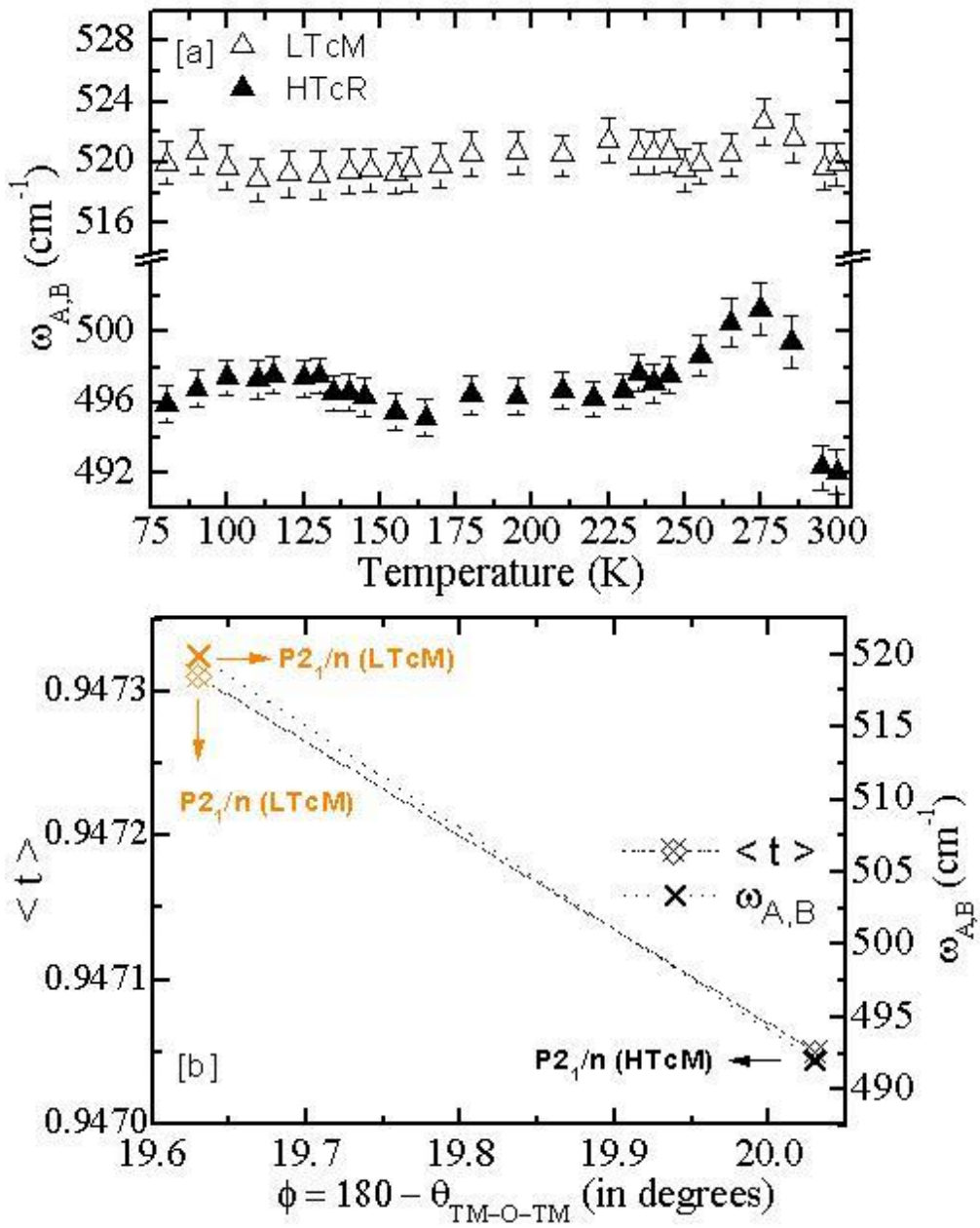

Figure 7. [a] Temperature dependence of $\omega_{A,B}$ for LTcM and HTcM. [b] Dependence of $\langle t \rangle$ and $\omega_{A,B}$ as function of octahedral tilt angle.

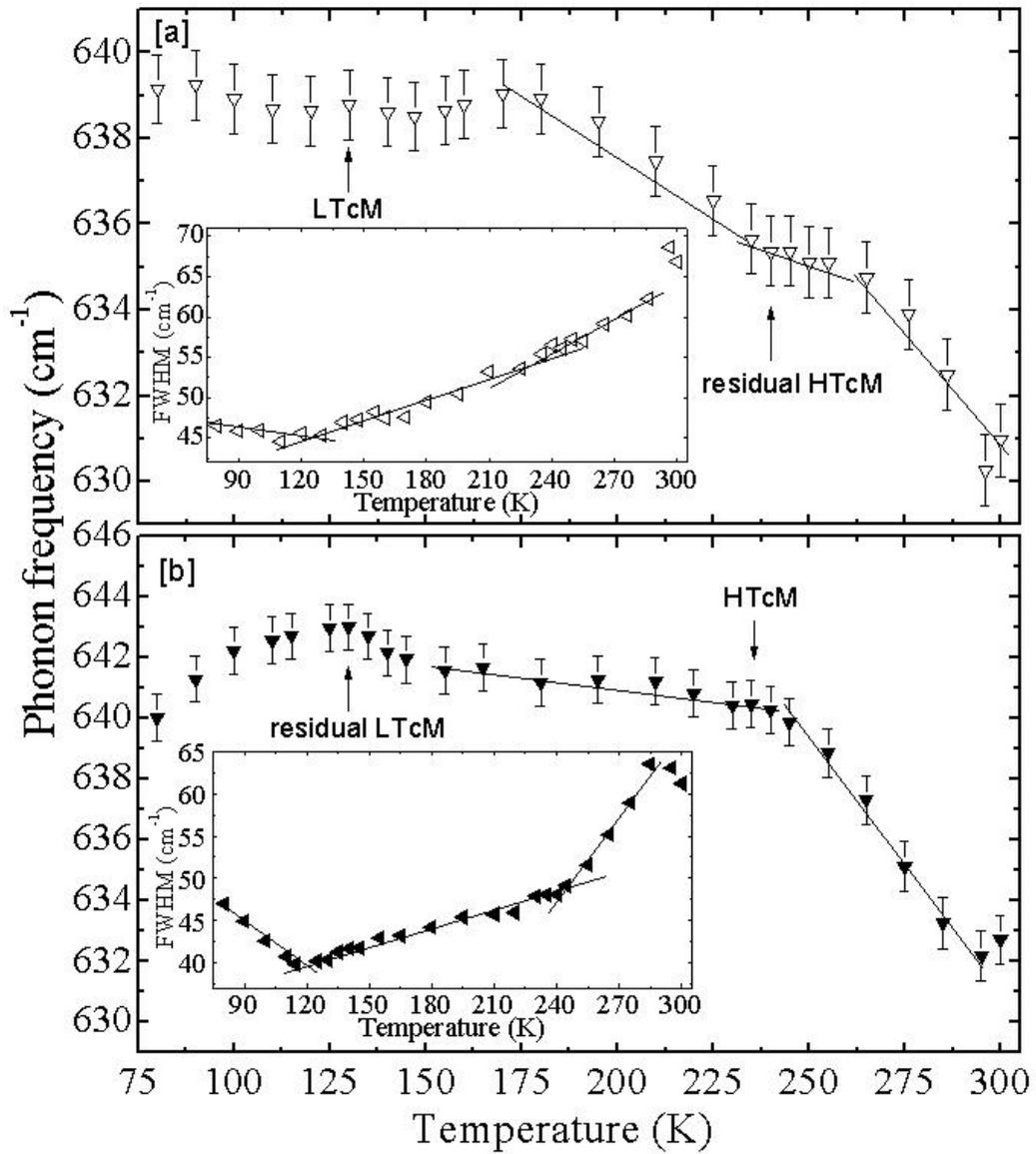

Figure 8. Temperature dependence of $\omega_S$, [a] LTcM and [b] HTcM; Inset showing their corresponding line width.

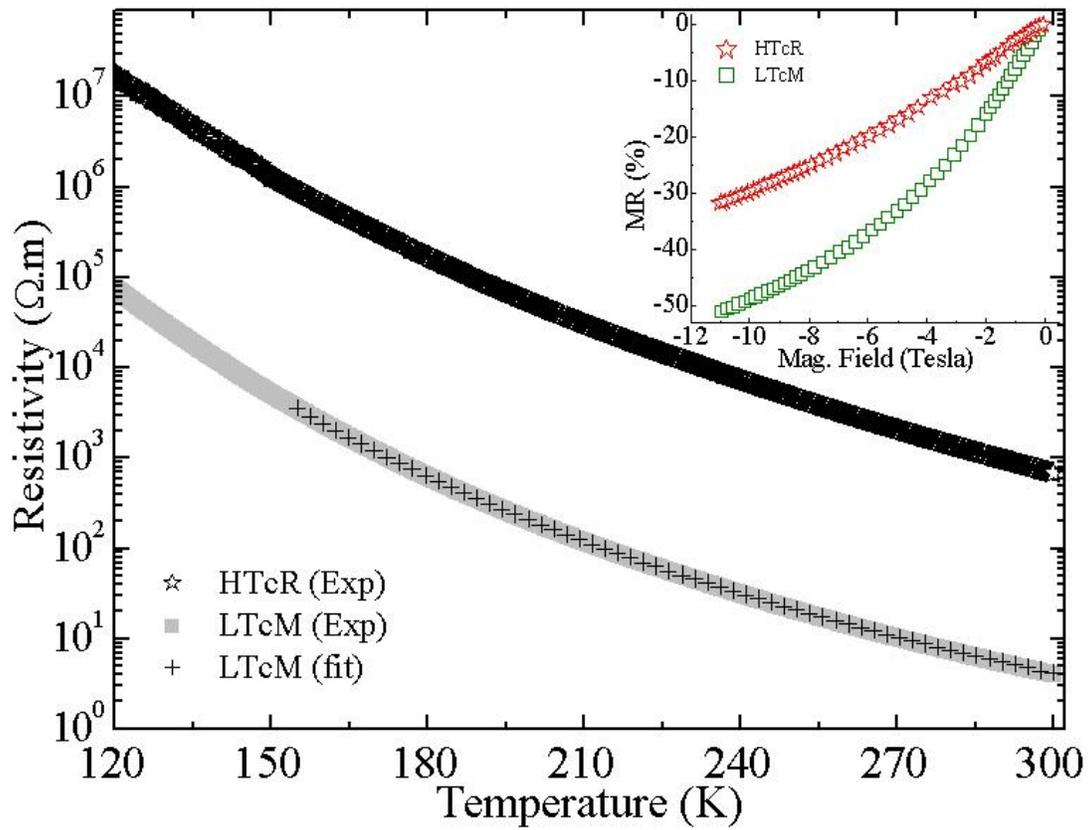

Figure 9. The temperature variation of resistivity of HTcR and LTcM phases. LTcM follows (ES-VRH) from 155 to 300 K; inset showing the Magnetoresistance (($R_{11T} - R_{0T}$) /$R_{0T}$) measured at 125 K for LTcM and HTcM.